\documentclass[aps,prl,reprint,superscriptaddress]{revtex4-2}
\usepackage{graphicx}
\usepackage{ragged2e}

\begin{document}


\title{One-Step Epitaxial Access to Rhombohedral Graphene Flat-Band States on Step-Bunched SiC}

\affiliation{Institute for Advanced Studies, Wuhan University, Wuhan 430072, China}
\affiliation{Department of Physics, Southern University of Science and Technology, Shenzhen 518055, China}
\affiliation{School of Physics and Technology, Wuhan University, Wuhan 430072, China}
\email{nxu@whu.edu.cn}
\author{Hao Zhong}
\affiliation{Institute for Advanced Studies, Wuhan University, Wuhan 430072, China}

\author{Xingzhe Wang}
\affiliation{Institute for Advanced Studies, Wuhan University, Wuhan 430072, China}

\author{Hanbin Deng}
\affiliation{Department of Physics, Southern University of Science and Technology, Shenzhen 518055, China}

\author{Tianyu Yang}
\affiliation{Department of Physics, Southern University of Science and Technology, Shenzhen 518055, China}

\author{Haixuan Cao}
\affiliation{School of Physics and Technology, Wuhan University, Wuhan 430072, China}

\author{Renzhe Li}
\affiliation{Institute for Advanced Studies, Wuhan University, Wuhan 430072, China}

\author{Qiang Wan}
\affiliation{Institute for Advanced Studies, Wuhan University, Wuhan 430072, China}

\author{Shangkun Mo}
\affiliation{Institute for Advanced Studies, Wuhan University, Wuhan 430072, China}

\author{Keming Zhao}
\affiliation{Institute for Advanced Studies, Wuhan University, Wuhan 430072, China}

\author{Shuming Yu}
\affiliation{Institute for Advanced Studies, Wuhan University, Wuhan 430072, China}

\author{Dingkun Qin}
\affiliation{Institute for Advanced Studies, Wuhan University, Wuhan 430072, China}

\author{Guang Zhu}
\affiliation{Institute for Advanced Studies, Wuhan University, Wuhan 430072, China}

\author{Yifan Zhou}
\affiliation{Institute for Advanced Studies, Wuhan University, Wuhan 430072, China}

\author{Jianping Shi}
\affiliation{Institute for Advanced Studies, Wuhan University, Wuhan 430072, China}

\author{Shuangfeng Jia}
\affiliation{School of Physics and Technology, Wuhan University, Wuhan 430072, China}

\author{He Zheng}
\affiliation{School of Physics and Technology, Wuhan University, Wuhan 430072, China}

\author{Jia-Xin Yin}
\affiliation{Department of Physics, Southern University of Science and Technology, Shenzhen 518055, China}

\author{Nan Xu}
\email{nxu@whu.edu.cn}
\affiliation{Institute for Advanced Studies, Wuhan University, Wuhan 430072, China}


\date{\today}

\begin{abstract}
Rhombohedral graphene multilayers provide a moir\'e-free platform for correlated and topological flat-band physics, but direct, transfer-free epitaxial access to thickness-tunable multilayers remains limited. Here we report a one-step graphitization route on 4$^\circ$ off-axis 4H-SiC, in which high-temperature flash annealing simultaneously drives self-organized step bunching and multilayer graphene formation. Atomic-resolution cross-sectional scanning transmission electron microscopy identify local ABC registry and distinguish rhombohedral from Bernal stacking. The thickness is tuned from bilayer to more than twenty layers by varying single parameter, the annealing temperature. Angle-resolved photoemission spectroscopy directly tracks the thickness-dependent evolution from interface-dominated low-energy states toward pronounced near-Fermi-level flat-band spectral weight in thick multilayers. Low-temperature scanning tunneling microscopy and spectroscopy on a 17-layer film further reveal a 13.4 meV low-energy spectral reconstruction and a $\sqrt{3} \times \sqrt{3}$ Kekul\'e-like modulation, providing microscopic signatures consistent with an intervalley-mixed electronic texture. This one-step, transfer-free approach establishes step-bunched SiC as an epitaxial platform that links stacking engineering with moir\'e-free correlated flat-band electronic states. 
\end{abstract}


\maketitle

Rhombohedral graphene multilayers (RGMs) provide a structurally simple and moiré-free platform for flat-band electronic physics. In ABC-stacked N-layer graphene, the low-energy chiral bands near charge neutrality follow an approximate dispersion $E \sim k^N$ \cite{MinMacDonald2008,KoshinoMcCann2009,Zhang2010}, becoming progressively flatter with increasing N and producing an enhanced density of states. For larger layer numbers, this evolution approaches a surface-flat-band regime \cite{Kopnin2011,HeikkilaVolovik2011}, associated with the Dirac-semimetal topology of rhombohedral graphite. The resulting high density of states makes rhombohedral multilayers susceptible to interaction-driven broken-symmetry phases, including correlated insulating, magnetic, ferroelectric, and layer-polarized states \cite{Shi2020,Kerelsky2021,Zhou2021Metals,Zhou2021Superconductivity,Han2023Multiferroicity,Han2024Chern,Liu2024,Winterer2024,Zhou2024Ferromagnetism}. Carrier density, displacement field, moiré potential, dielectric environment, and spin–orbit coupling have further been used to access Chern-insulating, quantum-anomalous-Hall, fractional-Chern, intervalley/isospin-related, and superconducting phases in rhombohedral tri-, tetra-, and pentalayer graphene \cite{Arp2024,Zhang2025Evolution,Auerbach2025,Han2024QAH,Sha2024,Lu2024FQAH,Xie2025,Patterson2025,Yang2025SOC,Han2025ChiralSC,Choi2025,Chatterjee2022,YouVishwanath2022}.

\begin{figure*}[!t]
{\centering
\includegraphics[width=0.84\textwidth]{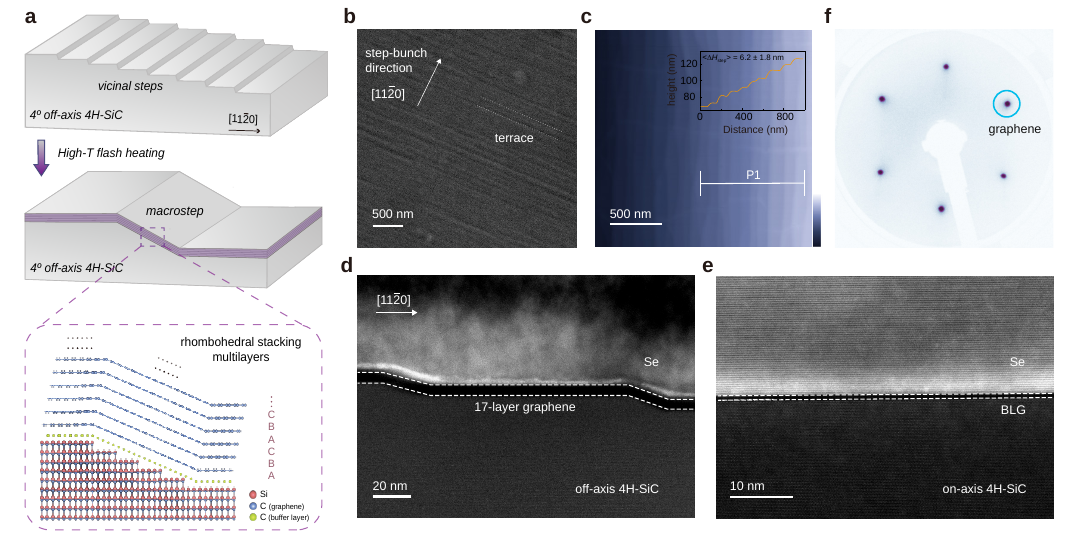}\par}

\caption{\justifying
Growth concept and morphology of step-bunched multilayer graphene on off-axis 4H-SiC.
(a) Schematic of flash annealing of 4$^\circ$ off-axis 4H-SiC, producing step bunching, macrosteps and multilayer graphene.
(b) SEM image of directional step-bunched terraces after annealing.
(c) AFM height map and line profile; the mean macrostep height is 6.19 $\pm$ 1.76 nm ($n = 36$, mean $\pm$ s.d.).
(d) Cross-sectional STEM image showing a conformal 17-layer graphene film on step-bunched off-axis SiC.
(e) Control STEM image of bilayer graphene grown on on-axis 4H-SiC under comparable conditions.
(f) LEED pattern acquired at 180 eV, showing ordered graphene diffraction features.
}
\label{fig:figure1}
\end{figure*}

Despite this progress, materials access remains a central bottleneck. Most high-quality devices rely on exfoliation and van der Waals assembly, which provide exceptional electronic quality but constrain sample size, yield, direct substrate compatibility, and systematic thickness control. Rhombohedral graphene has been reported through SiC graphitization, curvature-assisted stabilization, chemical-vapor-deposition and alloy-mediated routes \cite{NorimatsuKusunoki2010,Pierucci2015,Pierucci2016,Hajlaoui2016,Gao2020,Nguyen2020}. A recent independent study established step geometry–guided growth of high-purity rhombohedral graphene using an engineered stepped Al$_2$O$_3$/Cu--Ni architecture, followed by transfer and thermomechanical relaxation \cite{ZhaoEtAl2026}. That result demonstrates that crystallographically defined step boundaries can direct interlayer slip and stabilize rhombohedral stacking. It also sharpens the remaining materials question, whether the step template and rhombohedral multilayers can be generated simultaneously in a one-step, transfer-free process on the final crystalline substrate, while retaining direct access to their thickness-dependent and local correlated electronic states.

Here we report one-step epitaxial access to rhombohedral graphene multilayers on step-bunched 4° off-axis 4H-SiC. Flash annealing under ultrahigh vacuum simultaneously produces directional macrosteps and drives SiC graphitization, without an external carbon source, metal growth substrate, transfer, or post-growth mechanical reconstruction. Combining cross-sectional scanning transmission electron microscopy (STEM), angle-resolved photoemission spectroscopy (ARPES), and low-temperature scanning tunneling microscopy and spectroscopy (STM/STS), we identify local ABC registry, tune the thickness from bilayer to more than twenty layers. Furthermore, we directly observe near-Fermi-level flat-band spectral weight in the as-grown epitaxial films. STM/STS further reveals a low-energy spectral reconstruction and Kekulé-like intervalley modulation in a 17-layer film, establishing a direct connection among one-step growth, stacking registry, flat-band electronic structure, and candidate symmetry-broken electronic texture.

Figure 1a illustrates the self-organized, step-mediated graphitization process. The as-polished vicinal SiC substrate is expected to contain closely spaced atomic steps separated by relatively long terraces (Fig. 1a), although the intrinsic vicinal morphology is barely discernible before annealing. During high-temperature flash annealing under ultrahigh vacuum, these steps bunch into directional macrosteps while SiC simultaneously decomposes and supplies carbon for multilayer graphene growth. Si preferentially escapes from low-coordination step-edge, kink, and sidewall environments  \cite{Yazdi2016,Imoto2017,Sun2011,Kageshima2013}, whereas the residual carbon reconstructs at the carbon-rich SiC interface. Relative to flat terraces, which are rapidly passivated by a buffer layer or few-layer graphene, step-bunched regions provide additional Si-escape pathways and enhance local carbon accumulation. The same thermal process therefore creates both the growth template and the graphene multilayers.

\begin{figure}[t]
{\centering
\includegraphics[width=0.84\columnwidth]{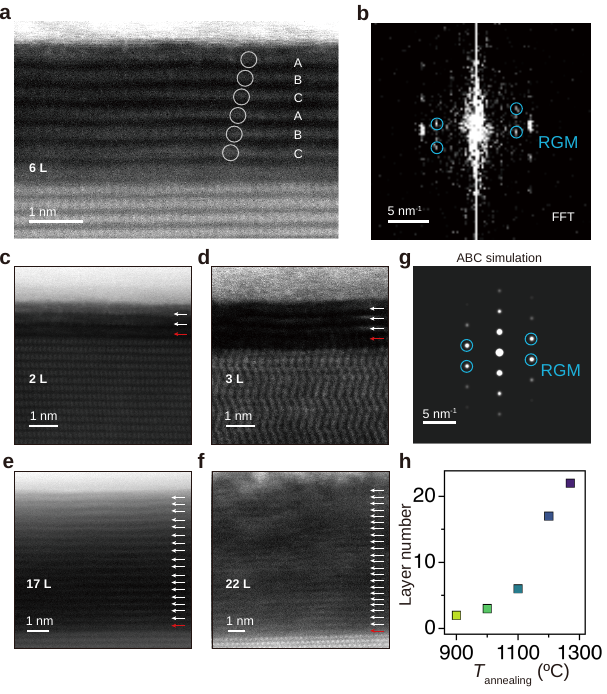}\par}

 \caption{\justifying
Rhombohedral stacking identification and annealing temperature dependent multilayer growth.
(a) Atomic-resolution cross-sectional STEM image of a representative six-layer region, showing ABCABC-like layer offsets.
(b) Corresponding FFT with rhombohedral-graphene features highlighted.
(c-f) Representative STEM images of samples grown at 900--1270 $^{\circ}\mathrm{C}$, with local layer numbers increasing from 2 to 22.
(g) Simulated diffraction patterns for the ABC model; cyan circles mark rhombohedral-characteristic reflections.
(h) Mean layer number versus annealing temperature.
}
\label{fig:figure2}
\end{figure}

Under a representative condition of flash annealing at $T_{\mathrm{a}} = 1200\,^{\circ}\mathrm{C}$, plan-view scanning electron microscopy reveals well-defined directional macrosteps on the off-axis SiC surface (Fig. 1b). Atomic-force-microscopy topography yields a mean macrostep height of 6.2 ± 1.7 nm (Fig. 1c). Cross-sectional STEM shows a conformal 17-layer graphene film across the macrostep region (Fig. 1d). In contrast, an on-axis 4H-SiC control treated under comparable graphitization conditions remains in the bilayer regime (Fig. 1e). This control establishes that ultrahigh-vacuum flash annealing alone is insufficient to account for thick multilayer formation and identifies the off-axis step-bunched morphology as a key growth variable. The LEED pattern contains ordered graphene diffraction features and strongly attenuated substrate reflections, consistent with substantial graphene coverage (Fig. 1f). Raman spectra provide complementary information on the multilayer graphitic character of the annealed films (Supplementary Fig. S1)

The present mechanism is distinct from the recently reported engineered step-confinement route [34]. There, a predefined stepped  Al$_2$O$_3$/Cu--Ni architecture controls the slip and orientation angles during carbon delivery through the metal substrate. Here, the macrostep template emerges spontaneously during the same Si-sublimation process that forms graphene on the final SiC substrate. Nevertheless, the two independent results support a broader principle, that crystallographically defined step geometries can couple interfacial growth to interlayer registry and provide a general route for accessing metastable rhombohedral stacking.

We next determine the local stacking sequence and its relation to thickness. Atomic-resolution cross-sectional STEM resolves the projected lateral registry of adjacent graphene sheets. In the representative six-layer region in Fig. 2a, the layer offsets follow an ABCABC sequence. Fourier analysis separates graphene-related reflections from the SiC contribution (Fig. 2b). The graphene thickness evolves systematically with flash-annealing temperature. Representative cross-sectional STEM images show graphene thicknesses of 2 and 3 layers at 900 and 1000 $^{\circ}\mathrm{C}$, respectively, increasing to 17 and 22 layers at 1200 and 1270 $^{\circ}\mathrm{C}$, respectively (Figs. 2c–f). A six-layer rhombohedral (ABCABC) structural model was constructed along the experimental projection direction, and the simulated diffraction pattern was compared under identical reciprocal-space calibration (Fig. 2g). The observed registry and the stacking-sensitive diffraction features agree with the rhombohedral model and distinguish it from the mirror-symmetric Bernal sequence. Additional  fast Fourier transform (FFT) and inverse-FFT analyses provide consistent real- and reciprocal-space signatures (Supplementary Fig. S2). Because these measurements probe spatially localized cross sections, we conservatively assign local rhombohedral-stacking signatures rather than a uniformly phase-pure rhombohedral film. The temperature-dependent layer-number statistics are summarized in Fig. 2h. Thus, the annealing temperature provides an effective parameter for tuning the representative local thickness on step-bunched SiC, from the few-layer regime to a thick multilayer limit while retaining direct epitaxial contact with the substrate.

\begin{figure}[t]
{\centering
\includegraphics[width=0.92\columnwidth]{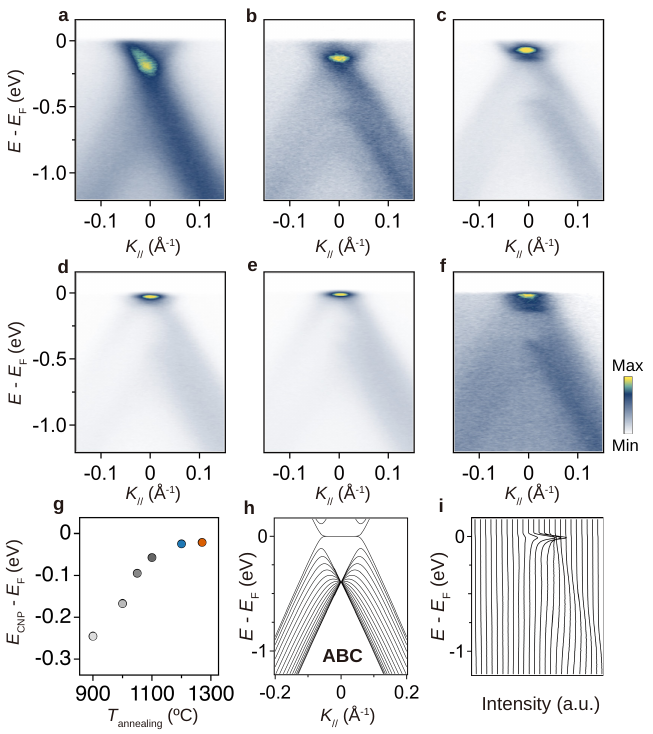}\par}
 \caption{\justifying
Layer-dependent low-energy electronic structure of epitaxial rhombohedral graphene multilayers.
(a-f)  ARPES intensity maps near the graphene K point for samples grown at 900--1270 $^{\circ}\mathrm{C}$, corresponding to representative local layer numbers of 2, 3, 5, 6, 17 and 22.
(g) Extracted $E_{\mathrm{CNP}} - E_{\mathrm{F}}$ as a function of annealing temperature, showing a shift of the charge-neutrality-point spectral weight towards the Fermi level.
(h) Calculated low-energy band structure of rhombohedral 17-layer graphene.
(i) Momentum-resolved EDCs extracted from \textbf{e} near the graphene K point.
}
\label{fig:figure3}
\end{figure}

This thickness series enables direct tracking of the low-energy electronic structure on the as-grown SiC platform. ARPES spectra acquired near the graphene K point display a pronounced thickness-dependent evolution of the charge-neutrality-point spectral weight (Figures 3a–f). In the bilayer sample grown at 900 $^{\circ}\mathrm{C}$, the dominant low-energy feature lies well below the Fermi level. With increasing annealing temperature and representative local thickness, the spectral weight shifts toward $E_{\mathrm{F}}$ and becomes concentrated near $E_{\mathrm{F}}$ in the 17- and 22-layer samples. The energy position $E_{\mathrm{CNP}}$, operationally defined from the dominant low-energy EDC feature near K, is summarized in Fig. 3g. Its systematic upward shift is consistent with progressive screening of interface-induced doping and a reduced relative contribution from the graphene/SiC interface in thicker multilayers \cite{Varchon2007,MattauschPankratov2007,Riedl2007,Emtsev2008,Nakatsuji2010}. Local strain, finite thickness variation, and stacking inhomogeneity may additionally affect the detailed line shape, particularly in the 22-layer sample.

Tight-binding calculations for 17-layer rhombohedral graphene yield pronounced surface-band spectral weight near charge neutrality (Fig. 3h). Consistent with this expectation, the experimental spectra develop an increasingly prominent, weakly dispersive near-$E_{\mathrm{F}}$ component with increasing thickness. The momentum-resolved EDCs of the 17-layer film follow the calculated rhombohedral surface-band profile (Fig. 3i), supporting assignment of this low-energy weight to rhombohedral flat-band states. A key advantage of our epitaxial SiC platform is that it enables direct tracking of the thickness-dependent electronic evolution across an as-grown sample series, establishing a continuous link among the growth conditions, representative local thickness, interface screening, and flat-band spectral weight without potential artifacts from transfer, mechanical thinning, or exfoliation.

The near-$E_{\mathrm{F}}$ flat-band spectral weight motivates a local search for interaction-sensitive electronic states. Low-temperature STM/STS measurements were therefore performed directly on an as-grown 17-layer film. Large-area topography reveals terraces extending over more than 100 nm, consistent with the step-bunched morphology (Fig. 4a). At 4.2 K, a representative dI/dV spectrum shows two pronounced low-energy density-of-states maxima separated by 13.4 meV and an intervening conductance suppression (Fig. 4b). This reconstruction occurs within the same energy range as the near-$E_{\mathrm{F}}$ spectral weight observed by ARPES.

Atomic-resolution topography and differential-conductance mapping further reveal a commensurate  $\sqrt{3} \times \sqrt{3}$ modulation with Kekul\'e character (Figs. 4c,d). The corresponding Fourier transforms contain inner peaks rotated by 30° relative to the graphene Bragg peaks (Figs. 4e,f), consistent with wave vectors connecting the $K$ and $K^{\prime}$ valleys. Bias-dependent conductance maps show that the inner peaks remain at essentially fixed reciprocal-space positions over the measured low-energy window, without a resolvable dispersive shift (Supplementary Figs. S3 and S4). The coexistence of a low-energy spectral reconstruction and a nondispersive Kekul\'e-like modulation provides microscopic signatures compatible with an intervalley-mixed or intervalley-coherent electronic texture in a thick, moir\'e-free rhombohedral multilayer. Importantly, the state is accessed directly on SiC without an engineered moiré superlattice or a spin–orbit-proximitized substrate.

The evidence should nevertheless be interpreted conservatively. STM/STS alone cannot uniquely distinguish a spontaneous intervalley-coherent ground state from structural reconstruction, local strain, substrate coupling, or defect-assisted intervalley scattering. The present data also do not establish a quantitative correspondence between the Kekulé peak intensity and the individual dI/dV maxima. Temperature dependence, spatial gap statistics, and correlation between the gap magnitude and Kekul\'e intensity would be valuable future tests of a genuine order parameter.

\begin{figure}[t]
{\centering
\includegraphics[width=0.9\columnwidth]{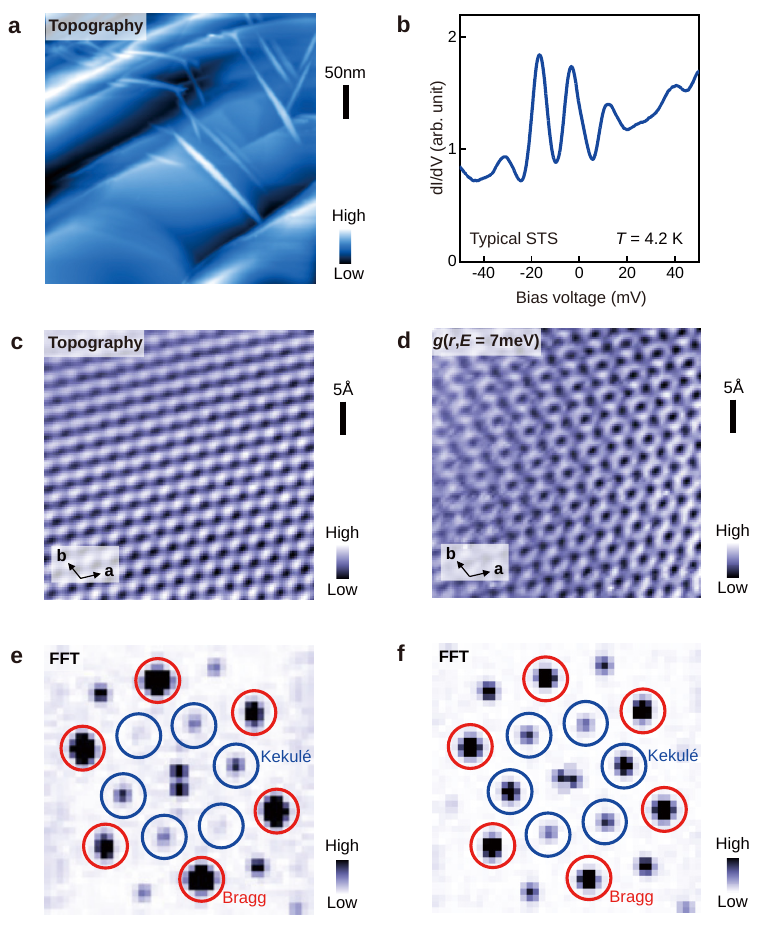}\par}
 \caption{\justifying
STM/STS measurements of a 17-layer epitaxial graphene film.
(a) Large-area STM topography showing terrace structures over a 400 $\times$ 400 nm$^2$ field of view.
(b) Representative $dI/dV$ spectrum acquired at 4.2 K, showing two low-energy density-of-states maxima separated by 13.4 meV.
(c,e) Atomic-resolution STM image and corresponding FFT. Red and blue circles mark graphene Bragg peaks and inner $\sqrt{3} \times \sqrt{3}$ peaks, respectively.
(d,f) Differential-conductance map acquired at $V_{\mathrm{bias}} = +7$ mV and corresponding FFT, showing inner peaks rotated by 30$^\circ$ relative to the graphene Bragg peaks, consistent with a Kekul\'e-like $\sqrt{3} \times \sqrt{3}$ modulation.
}
\label{fig:figure4}
\end{figure}

\section*{Conclusion}

We have established a one-step, transfer-free route to rhombohedral graphene multilayers on step-bunched 4° off-axis 4H-SiC. The same flash-annealing process simultaneously generates directional macrosteps and drives interfacial graphitization. Local ABCABC registry is resolved by atomic-scale STEM, FFT analysis, and diffraction simulations, while the representative local thickness is tuned from bilayer to more than twenty layers through the annealing temperature. Direct ARPES measurements on the as-grown films reveal progressive emergence of near-Fermi-level rhombohedral flat-band spectral weight. STM/STS on a 17-layer film further identifies a 13.4 meV low-energy spectral reconstruction and a nondispersive Kekulé-like intervalley modulation. Together, these results connect self-organized step engineering, stacking registry, thickness control, and correlated low-energy electronic structure within a single epitaxial SiC platform.

\section*{Methods}

Graphene multilayers were grown by flash annealing commercial 4° off-axis 4H-SiC substrates under ultrahigh vacuum at 900–1270 $^{\circ}\mathrm{C}$. Surface morphology was characterized by SEM and AFM, surface crystallinity by LEED, and layer number and stacking registry by cross-sectional STEM, with additional characterization by Raman spectroscopy. The low-energy electronic structure was investigated by ARPES and low-temperature STM/STS and compared with tight-binding calculations. Full experimental details are provided in the Supplementary Information.

\section{ASSOCIATED CONTENT}

The Supporting Information Available:
Additional growth parameters and temperature calibration; Raman characterization; additional ARPES measurements; bias-dependent STM/STS conductance maps; and the corresponding FFT analyses.

\section{ACKNOWLEDGMENTS}
The authors acknowledge the Center for Electron Microscopy and the Core Facility of Wuhan University for their support with electron microscopy and surface characterization. We thank Dr. Ying Zhang, Dr. Wenhua Xue, and Dr. Ye Ma from the Core Facility of Wuhan University for their assistance with FIB/SEM, TEM, and AFM analyses, respectively. We also thank the group of Prof. Jianping Shi for assistance with Raman measurements performed using an XploRA PLUS Raman microscope.

\clearpage
\bibliography{references}

\end{document}